\documentstyle[prl,aps,epsfig,amssymb]{revtex} 
 
\begin{document} 
\twocolumn[
\hsize\textwidth\columnwidth\hsize\csname@twocolumnfalse\endcsname
\title{The Spin Glass Transition : Exponents and Dynamics} 
 
\author{L.W. Bernardi, N. Lemke, P.O. Mari and I.A. Campbell} 
\address{Physique des Solides, Universit\'{e} Paris Sud, 91405  
Orsay, France} 
\author{A. Alegr\'{\i}a and J. Colmenero}
\address{Departamento de F\'{\i}sica de Materiales, \\Facultad de
Qu\'{\i}mica, UPV/EHU, Apartado 1072, 20080 San Sebasti\'an, Spain}

\maketitle

\begin{abstract} 
Numerical simulations on Ising Spin Glasses show that spin glass  
transitions do not obey the usual universality rules which hold at canonical  
second order transitions. On the other hand the dynamics at the approach  
to the transition appear to take up a universal form for all spin glasses. The  
implications for the fundamental physics of transitions in complex systems  
is addressed. 
\end{abstract} 
\pacs
\twocolumn ]
\narrowtext
 
  Spin glasses are magnetic systems with conflicting and random interactions  
between individual spins. The denomination "spin glass" for the wide class  
of such materials which show a complex form of magnetic order at low  
temperatures dates from 1970 \cite{ref1}; the same year and independently  
it was found that a dilute magnetic alloy has a sharp cusp in the low field  
a.c. susceptibility at a temperature $T_g$\cite{ref2}, so defining a precise  
critical temperature despite the disorder. Spin glasses have been intensively  
studied ever since, theoretically, experimentally and numerically in the belief  
that they will provide a paradigm for a vast family of complex systems. 
 
Numerical work has concentrated on the Edwards-Anderson version of a  
spin glass -- a regular lattice of Ising spins coupled by random interactions.  
It turns out that this "theoretician's spin glass" shows the same essential  
characteristics as the experimental systems, and is a perfect test-bed for  
checking out ideas on complex systems in general.  The mean field (or  
infinite dimension) limit of this model is well understood \cite{ref3}.  
However, in finite dimensions even this simplified version of a complex  
system turns out to be very devious, so that  after 25 years of work there is  
still little concensus on the fundamental empirical facts, let alone the  
underlying theory. 
 
At low temperatures there is a domain of irreversible, non-ergodic  
behaviour; we will discuss exclusively the dynamics in the other, ergodic,   
regime from high temperatures down to (or nearly down to) the freezing  
temperature $T_g$. This critical temperature can be defined operationally  
by the divergence of the relaxation time for large systems, or by invoking  
scaling laws.  It is well known that at a standard second order transition one  
can define critical exponents which are related to each other and which have  
been shown, by the renormalization group theory, to obey universality rules.  
These state essentially that in a given dimension, all systems sharing a few  
basic parameters (such as the number of order parameter components) will  
have identical values of critical exponents. Physically this remarkable  
regularity arises because averaging over larger and larger volumes as the  
critical temperature is approached means that all microscopic distinctions  
between systems are blurred out. 
 
There have been a number of numerical estimates of critical exponents in  
spin glasses. A first point which is clear is that the estimated exponents at a  
spin glass transition are very different from those at a standard second  
order transition. For instance, in dimension 3 the specific heat exponent  
$\alpha$ for an Ising ferromagnet is close to zero; in the Ising spin glass it is  
close to $-2$ \cite{ref4,ref5}. It is then important to check if the universality  
rules hold at spin glass transitions as they do at canonical second order  
transitions. This turns out to be difficult to do in practice, mainly because  
the relaxation as the transition is approached becomes exceedingly slow,  
meaning that long and tedious numerical anneals must be used to make sure  
that the system is in equilibrium. As a result, simulation estimates of  
exponents have had substantial error bars associated with them.  
 
We have used a combination of scaling rules which allow us in favourable  
cases to establish accurate values of the ordering temperatures together with  
the static exponent $\eta$ and the dynamic exponent $z$ \cite{ref6,ref7}.  
Among these techniques is that of the scaling form of the non-equilibrium  
relaxation in going from complete disorder towards equilibrium, after a  
numerical quench to the critical temperature. This has the advantage of  
needing no preparatory anneal. It would be much more elegant if it were  
possible to establish the ordering temperature uniquely from such non- 
equilibrium measurements, as has been done in regular magnetic systems  
\cite{ref8}. Unfortunately up to now we have found no simple method of  
doing this.  
 
The exponent values that we have estimated are more precise than but in  
good agreement with other simulation estimates \cite{ref4,ref5}, and with  
exponent values obtained from series methods \cite{ref9}. The results  
clearly indicate that the exponents $\eta$ and $z$ for spin glasses in a given  
dimension change with the form of the random interactions. For instance,  
the exponents found with the binomial near neighbour interaction  
distribution $\pm J$ are not the same as those found with a Gaussian near  
neighbour interaction distribution. According to the universality rules, such  
differences in the interactions should not be pertinent. One would expect,  
with a change in distribution, a change in the critical temperature but not a  
change in the exponents. We can note that in experimental spin glass  
systems the estimates for $\eta$ vary considerably from one material to  
another, again suggesting a breakdown of universality \cite{ref10}. It seems  
that ordering at a spin glass transition cannot be simply classified with  
standard second order transitions. This is an important point, as the general  
renormalization argument would seem to be very robust so the breakdown  
is unexpected and goes against the conventional wisdom.  The empirically  
observed non-universality implies that the physics of critical behaviour in  
spin glasses and other complex systems is much richer than in regular  
systems. It would be very useful to find a theoretical underpinning for this  
result, and to understand how the exponents change with the various  
control parameters such as the interaction distribution. 
 
As we have noted above, relaxation at the approach to a spin glass  
transition becomes extremely slow. It also becomes highly non-exponential.  
In a massive simulation Ogielski \cite{ref4} measured accurately the  
relaxation of the autocorrelation function  
$q(t)=<S_i(t).S_i(0)>$ 
for large samples of the $\pm J$ Ising spin glass in dimension 3. At the  
approach to a continuous transition one expects $q(t)$ to take the form  
$t^{-x} f(t/\tau)$ where $x$ is a combination of critical exponents and  
$\tau$ is a characteristic time which diverges at the critical temperature.  
Ogielski parametrised his results very succesfully using for the scaling  
function $f$ the stretched exponential form, $exp[-(t/\tau)^{\beta}]$, with a  
temperature dependent stretched exponential (or Kohlrausch \cite{ref11})  
exponent $\beta$. The stretched exponential is used ubiquitously by the  
glass community for fitting experimental relaxation data. Having $\beta$ less  
than 1 can be interpreted as an indication of a distribution of relaxation  
times in the system. In Ogielski's data, as the temperature approaches  
$T_g$ and $\tau$ diverges, $\beta$ tends to a value close to 1/3. Numerical  
data on different spin glasses are compatible with similar behaviour in each  
of the other cases \cite{ref12,ref13}. However good the data, it is of course  
always necessary to extrapolate to estimate the limiting value of $\beta$ as  
$\tau$ diverges, which leads to a margin of uncertainty. Nevertheles the  
general tendency appears to be very similar in all the spin glasses which  
have been studied.  
 
In experimental spin glass systems it is difficult to study the form of the  
relaxation for a range of temperatures  above $T_g$ directly because of  
time scale problems, but muons can be used as local probes which are  
sensitive to the relaxation of the magnetic spins. The form of the decay of  
the muon polarization with time gives indirect information on the form of the  
spin relaxation, and it turns out that the muon time window is appropriate.  
Measurements have shown that in a standard metallic spin glass the spin  
relaxation is homogeneous and close to exponential only at temperatures of  
the order of 4 times $T_g$, and that the muon polarization decay pattern is  
compatible with an Ogielski form of spin relaxation for temperatures close  
to $T_g$\cite{ref14,ref15}. This means that the behaviour observed  
numerically in Ising spin glass simulations is mirrored in experimental,  
Heisenberg, spin glasses. 
 
This pattern of behaviour is in fact even more general and is not confined to  
spin glasses alone. The relaxation in most structural glasses has a  
characteristic time which appears to diverge at a temperature $T_0$  
somewhat below the "glass temperature" $T_g$. (In the structural glass  
context the latter is conventinally defined as a temperature where the  
relaxation time scale is of the order of a few seconds. This is a convenient  
parameter for defining the practical onset of glassy as against liquid  
behaviour, but this glass $T_g$ is obviously not a thermodynamic  
temperature. $T_0$, the divergence temperature, is the equivalent of the  
spin glass freezing temperature).  The relaxation data are well fitted by  
stretched exponentials, and for a large number of glasses $\beta (T)$ drops  
regularly as $T_0$ is approached, with $\beta$ tending to a value in the  
region close to 1/3 when the points are extrapolated to the temperature  
where the relaxation time becomes infinite \cite{ref16,ref17,ref18}. Once  
again, extrapolation is obligatory, as the time needed to establish equilibrium  
before the start of the relaxation measurement excedes laboratory time  
scales well before $T_0$ is attained. Nevertheless there is a striking  
similarity between the behaviour of the relaxation in the Ising spin glasses  
and in these structural glasses, two families of systems which have nothing  
in common on the microscopic level. This suggests that in these systems  
the special pattern of the dynamics as the freezing temperature is  
approached is a necessary consequence of the physics of the glassy  
transition. 
 
We have proposed a possible interpretation couched in terms of the  
morphology of phase space at temperatures above the glass transition  
\cite{ref19,ref20}. Let us concentrate on the Ising systems. The total phase  
space of an N-spin Ising system consists of $2^N$ configurations which  
can be mapped onto the $2^N$ corners of a hypercube of dimension N. At  
finite temperature $T$ some of these configurations have such high energies  
that they are thermodynamically "inaccessible". The system performs a  
random walk among the low energy  configurations that remain  
"accessible". Relaxation through successive single spin flips is simply a  
random walk where each step takes the system from one configuration to a  
near neighbour configuration on the hypercube. The difficult problem is to  
establish the morphology of the set of accessible configurations, given the  
Hamiltonian for the system. Suppose that in a glassy system we take an  
extreme limit and assume that a temperature $T$ corresponds to a fraction  
$p$ of accessible sites, distributed at random over the hypercube. The  
relaxation problem is now mapped onto a geometrical problem. As $p$  
drops we will have the closed space equivalent of a percolation transition,  
with a single giant cluster of sites existing down to a critical concentration  
$p_c$ \cite{ref21}. Relaxation becomes a random walk among the sites of  
this giant cluster. We have argued \cite{ref19}, using the known behaviour  
of random walks on a percolation cluster in Euclidean space \cite{ref22},  
that for the high dimension hypercube relaxation will approach the stretched  
exponential form $exp[-(t/\tau)^1/3)]$ as $p_c$ is approached. Numerical  
simulations on this geometrical model confirm this  
conjecture\cite{ref19,ref20}.  
 
Our explanation of the particular form of the relaxation as $\tau$ diverges at  
a glass transition is then the following. At the transition the accessible phase  
space splinters into many topologically disconnected and inequivalent  
pieces, like the small clusters below the critical concentration in a  
conventional percolation transition. At a temperature just above the critical  
temperature the accessible phase space is still topologically connected, but  
it has become so sparse and complicated (fractal-like) that the system takes  
a macroscopic time to explore all the nooks and crannies of this structure,  
in order to be able to make up its mind whether the attainable phase space is  
indeed topologically connected or not. As the random walk path is  
tortuous, the relaxation is strongly non-exponential, and the hypercube  
geometry argument leads us to a specific form of relaxation. We have  
discussed the Ising phase space explicitly, but any physical system has a  
phase space and the general image should carry over to other glassy  
transitions. This shattering of phase space at the transition is very different  
from the description of a second order transition in a regular system where  
at the ordering temperature phase space splits up into two (or a small  
number) of configuration clusters related to each other by symmetry. One  
can surmise that the non-universality of the critical exponents is also a  
consequence of the geometry of phase space, but the connection remains  
to be established. 
 
We should at this point make a caveat. The discussion as it stands relates to  
glasses which are known as "fragile" glasses in the glass terminology, i.e. to  
glasses where $T_0$ is relatively close to $T_g$. In "strong" glasses,  
where the relaxation time grows following an Arrhenius law as the  
temperature is reduced to $T_g$, it appears to be the elementary step time  
which has increased to a macroscopic value at $T_g$, and it is impossible  
to probe close to $T_0$ (supposing it exists) in these glasses. $\beta$  
remains close to $1$ down to $T_g$ for these cases, which is entirely  
consistent with the general arguments we have given. In intermediate cases  
the extrapolation down to $T_0$ can only be made with a considerable  
degree of uncertainty. Relating the degree of fragility to the molecular  
structure of each glass remains a fundamental problem. 
 
In conclusion, we have discussed numerical and experimental data on spin  
glasses and glasses. The results indicate empirically that the conventional  
universality rules are not obeyed at spin glass transitions. On the other hand,  
as the critical temperature (where the relaxation time scale diverges) is  
approached, the form of the relaxation in spin glasses or fragile structural  
glasses tends towards a stretched exponential with a Kohlrausch exponent  
close to 1/3. We have provided a general phase space image that could  
explain these observations. 
 
\section {acknowledgements} 
Much of this work had the support of a COFECUB collaboration.  
Numerical calculations could be made thanks to a time allocation provided  
by IDRIS ( Institut du D\'{e}veloppement des Ressources en Informatique  
Scientifique). N.L. was supported by a CNPq scholarship.

\begin{figure}
\begin{center}
\epsfig{file=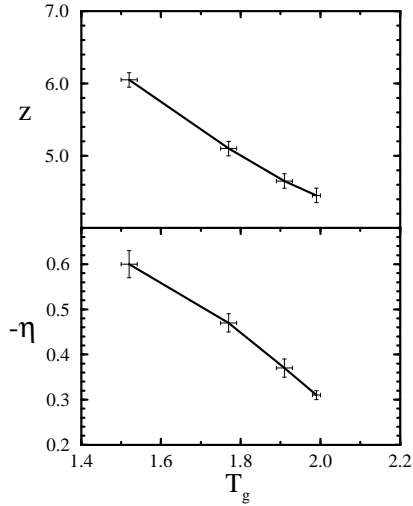,width=8.75cm}
\end{center}
\caption{The dynamic exponent $z$ and the critical exponent $\eta$ as 
functions of the critical temperature $T_g$ for four Ising spin glasses
in dimension 4. The sets of interactions, from left to right, are 
exponential, gaussian, uniform and binomial \protect\cite{ref7}.}
\label{figure:1}
\end{figure}

\begin{figure}
\begin{center}
\epsfig{file=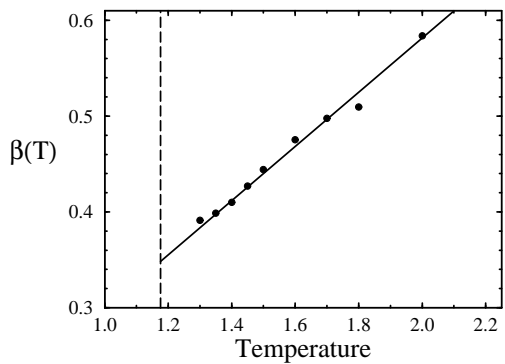,width=7.0cm,bbllx=18,bblly=-1,bburx=552,bbury=440}
\end{center}
\caption{The Kohlrausch exponent $\beta$ for the relaxation of the 
autocorrelation function as a function of $T$, for the 3D Ising spin 
glass model with binomial interaction distribution \protect\cite{ref4}. 
The points extrapolate to close to $1/3$ at the critical temperature
$T_g \simeq 1.175J$ (dashed line).}
\label{figure:2}
\end{figure}

\begin{figure}
\begin{center}
\epsfig{file=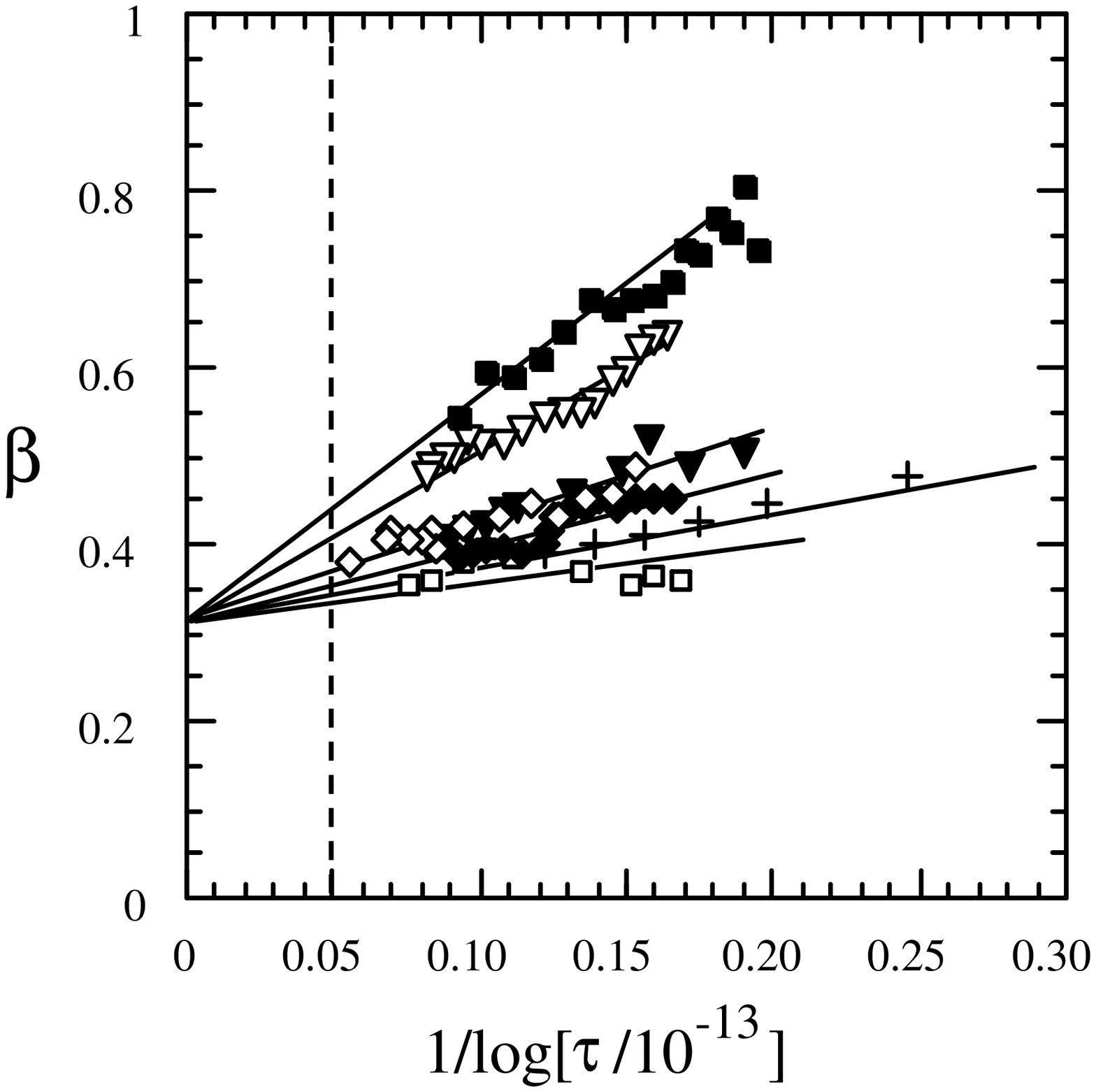,width=8.4cm,bbllx=15,bblly=175,bburx=580,bbury=750}
\end{center}
\caption{The Kohlrausch exponent $\beta(T)$ as a function of 
$(log(\tau(T)10^{13}))^{-1}$ obtained from dielectric relaxation measurements
\protect\cite{ref17} on a selection of polymers (PSF ($\blacksquare$), 
PViBE($\triangledown$), PE ($\lozenge$), PEI ($\blacktriangledown$), 
PH ($\blacklozenge$) and PMA ($\square$)) and on the 3D Ising spin 
glass model with binomial interaction distribution ($+$)\protect\cite{ref4}.
The dashed line represents the practical limit ($\tau \sim 4$ months) for 
equilibrium measurements. All the sets of data are compatible with an 
extrapolated value of $\beta$ near $1/3$ as $\tau$ goes to infinity (the 
lines through $\beta= 1/3$ are to guide the eye).}
\label{figure:3}
\end{figure}

\end{document}